\documentclass[prd,aps,lengthcheck,twocolumn,nofootinbib,notitlepage,floatfix,superscriptaddress]{revtex4-2}
\usepackage{graphics,graphicx}
\usepackage{amsmath, amssymb}
\usepackage{multirow}
\usepackage{color}
\usepackage{verbatim}
\usepackage{hyperref}
\usepackage[normalem]{ulem}
\usepackage{ulem}
\usepackage{color}
\usepackage{cancel}
\usepackage{mathtools}

\def\fm3{\;\text{fm}^{-3}}

\def\nsat{n_{\rm sat}}
\def\xeft{{\chi \rm EFT}}
\def\pqcd{{\rm pQCD}}

\def\nkt{n_{\rm th}}
\def\ukt{\mu_{\rm th}}
\def\pkt{p_{\rm th}}

\renewcommand\sout{\bgroup\color{blue} \ULdepth=-.5ex \ULset}
\def\slashchar#1{\setbox0=\hbox{$#1$}  % set a box for #1
\dimen0=\wd0     % and get its size
\setbox1=\hbox{/} \dimen1=\wd1  % get size of /
\ifdim\dimen0>\dimen1   % #1 is bigger
\rlap{\hbox to \dimen0{\hfil/\hfil}} % so center / in box
#1     % and print #1
\else     % / is bigger
\rlap{\hbox to \dimen1{\hfil$#1$\hfil}} % so center #1
/      % and print /
\fi}

\newcommand{\cs}{c_s^2}

\newcommand{\eps}{\epsilon}
\newcommand{\dd}{\mathrm{d}}

\begin{document}
% \title{quarkyonic matter and baryquark matter}
% \title{Dynamical generation of momentum-space shell and the chiral symmetry restoration}
\title{Kinetic-Theory Bounds on the Equation of State of Dense QCD Matter}

\author{Micha\l{} Marczenko}
\email{michal.marczenko@uwr.edu.pl}
\affiliation{Institute of Theoretical Physics, 
University of Wroc\l{}aw, plac Maksa Borna 9, 50-204 Wroc\l{}aw, Poland}
\affiliation{Incubator of Scientific Excellence - Centre for Simulations of Superdense Fluids,
University of Wroc\l{}aw, plac Maksa Borna 9, 50-204 Wroc\l{}aw, Poland}

\date{\today}

\begin{abstract}
We derive bounds on the equation of state of cold, dense matter by extending the causal, model-agnostic interpolation between chiral effective field theory and perturbative calculations with a microscopic constraint from relativistic kinetic theory. The additional condition restricts the stiffest admissible behavior of the equation of state and systematically reduces the range of allowed equations of state, with the strongest effect at high densities. The resulting bounds remain consistent with known low- and high-density limits, while the strength of the constraint depends on the density above which the kinetic-theory condition is applied. These bounds can be readily incorporated into future studies of cold, dense matter and used to assess the impact of microscopic stability conditions on equation-of-state inference.
\end{abstract}

\maketitle

\section{Introduction}

Recent advances in neutron-star (NS) astronomy, from precise radius measurements~\citep{Miller:2021qha, Riley:2021pdl} and the discovery of massive pulsars~\citep{Demorest:2010bx, Antoniadis:2013pzd} to gravitational-wave detections of binary mergers~\citep{LIGOScientific:2017vwq, LIGOScientific:2017ync}, have opened direct access to strongly interacting matter at supranuclear densities. Within general relativity, the macroscopic structure of NSs is governed by the equation of state (EOS) of cold and dense quantum chromodynamics (QCD) matter. Although, in principle, the EOS is uniquely determined by QCD, reliable first-principle calculations exist only in asymptotic limits. At low densities, chiral effective field theory ($\xeft$) provides a systematic expansion valid up to around nuclear saturation density $\nsat$~\citep{Tews:2018kmu, Hebeler:2013nza, Drischler:2017wtt, Drischler:2020fvz, Drischler:2020hwi, Keller:2022crb, Drischler:2020yad}, while at very high densities, perturbative QCD ($\pqcd$) becomes applicable for $n \gtrsim 40\,\nsat$~\citep{Hebeler:2013nza, Tews:2018kmu, Gorda:2021znl, Gorda:2021gha}. The intermediate regime realized in NS cores, corresponding to strongly coupled QCD matter at a few times $\nsat$, remains inaccessible to controlled {\it ab~initio} calculations and at present must be constrained indirectly~(see, e.g.,~\cite{Ecker:2022dlg, Ecker:2022xxj, Annala:2017llu, Annala:2019puf, Marczenko:2022jhl, Marczenko:2023txe, Brandes:2023hma}). 

A model-independent interpolation between the $\xeft$ and $\pqcd$ regimes provides a systematic way to connect nuclear and quark matter equations of state under basic thermodynamic consistency~\citep{Komoltsev:2021jzg}. This approach enforces the most fundamental physical principles, thermodynamic stability, causality, and integrability, without relying on any specific microscopic model. It defines an admissible region in the $(n,\mu)$ plane whose boundaries correspond to the stiffest and softest EOSs consistent with these requirements. When mapped to the $(\eps,p)$ plane, the resulting family of functions forms a band that encompasses all physically allowed equations of state of dense QCD matter. The approach has attracted growing interest and is now commonly used in the EOS analyses (see, e.g.,~\cite{Fujimoto:2023unl, Gao:2025ncu, Alarcon:2025qmz}).

The macroscopic causality condition $c_s^2 \le 1$ constrains only the propagation of signals. The stability of Israel-Stewart (IS) theory further restricts the admissible dynamics. The relaxation times of dissipative currents must be within certain limits, and if kinetic theory (KT) is applied to evaluate them, then within this approximation, one obtains a stronger inequality for the speed of sound~\citep{Olson:2000vx},
\begin{equation}\label{eq:cs2_kinetic_eq}
c_{s}^2 \leq \frac{\eps - p/3}{\eps + p} \equiv c_{s,\, \rm KT}^2 \textrm,
\end{equation}
where $p$ is the pressure and $\eps$ is the energy density. The limit is derived from the requirement that the relaxation time in a causal Boltzmann equation remains positive. At low densities, where $p \ll \eps$, this condition reproduces the causal bound $c_s^2 \lesssim 1$, while in the conformal limit, where $p = \eps/3$, it gives $c_s^2 \le 2/3$, which safely accommodates the conformal value $c_s^2 = 1/3$. Thus, the KT bound restricts the maximal stiffening of any EOS that admits a quasi-particle transport description. It has been examined in studies of dissipative hydrodynamics and applied phenomenologically to model NS EOSs and constrain the sound speed in dense QCD matter (see, e.g.~\cite{Moustakidis:2016sab, Margaritis:2019hfq, Laskos-Patkos:2024otk}).

In this work, we extend the causal-integrable construction of~\cite{Komoltsev:2021jzg} by incorporating the KT bound as a microscopic constraint on the EOS. The onset density $\nkt$ (or equivalently, chemical potential $\ukt$) above which the KT condition is enforced is treated as a free parameter, reflecting the uncertainty in the regime where kinetic theory becomes applicable (see, e.g.,~\cite{Denicol:2012cn, Romatschke:2017ejr} for discussion of the applicability of kinetic theory and Israel-Stewart truncations). We show that the KT bound substantially narrows the EOS band relative to the purely causal case, illustrating how transport-theory stability conditions translate into quantitative restrictions on the high-density equation of state.

% kinetic theory here refers to the broad class of transport descriptions based on a Boltzmann (or Boltzmann-like) equation and is, in principle, more generally applicable than any particular truncation scheme such as the Israel–Stewart (IS) 14-moment approximation.  In practice, however, quantitative results depend on (i) the dominant microscopic degrees of freedom, (ii) the strength and nature of interactions, and (iii) the closure/truncation used to derive fluid-dynamic equations and estimate relaxation times.  Consequently, we adopt the conservative strategy of treating the KT prescription as an \emph{effective} transport-inspired constraint: it indicates the direction and scale of microphysical limitations on stiffness but is not asserted as a controlled QCD result at all densities.  This motivates our operational choice to impose the KT inequality only above a variable onset density $\nkt$, and to present results as a function of $\nkt$ so that the reader can assess the sensitivity of our bounds to the assumed regime of transport applicability.  

% \section{Bounds on baryon number density}
\section{Bounds from causality}
\label{sec:causal}

In this work, we consider the construction of all possible interpolations of the thermodynamic pressure $p(\mu)$ at zero temperature between two known values of the chemical potential, $\mu_L$ and $\mu_H$ (see~\cite{Komoltsev:2021jzg} for details). We assume that the pressures $p(\mu_L)$, $p(\mu_H)$ and baryon densities $n(\mu_L)$, $n(\mu_H)$ are also known. Here, we use the constraints from $\xeft$ and pQCD to fix the EOS at $\mu_L$ and $\mu_H$, respectively. The variation of the low-density EOS within the $\chi$EFT constraint does not lead to qualitative changes in the results presented in this work~\citep{Komoltsev:2021jzg}. Therefore, we use a single EOS from $\xeft$, corresponding to the EOS with the largest pressure at $\mu_{\xeft} = 0.978\,$GeV, $p_\xeft = 3.542\,\rm MeV/fm^3$, and $n_\xeft=0.176\,\rm fm^{-3}$~\citep{Hebeler:2013nza}. On the other hand, the high-density pQCD constraint is renormalization-scale dependent, and variation of the scale parameter $X$ has to be taken into account. Following~\cite{Komoltsev:2021jzg}, we fix the pQCD constraint at $\mu_H=2.6\,$GeV and use $p_{\pqcd} = 2.334,\,3.823,\,4.284\,\rm GeV/fm^3$ and $n_{\pqcd} = 6.14,\,6.47,\,6.87\,\rm fm^{-3}$ for $X=1,\,2,\,4$, respectively.

To constrain the allowed family of densities, $n(u)$, we use the conditions of thermodynamic consistency and require stability and causality, which together ensure that $0 \leq c_s^2 \leq 1$. The first inequality implies that the density $n(\mu)$ is a non-decreasing function. The second inequality implies that
\begin{equation}
    c_s^2 = \frac{n}{\mu}\frac{\dd \mu}{\dd n} \leq 1 \; \Longrightarrow \; \frac{\dd n}{\dd \mu} \geq \frac{n}{\mu} \textrm,
\end{equation}
which determines the smallest possible slope of $n(\mu)$, i.e., the stiffest allowed EOS at any given point $(\mu,n)$.

\begin{figure}[t!]
    \centering
    \includegraphics[width=1.0\linewidth]{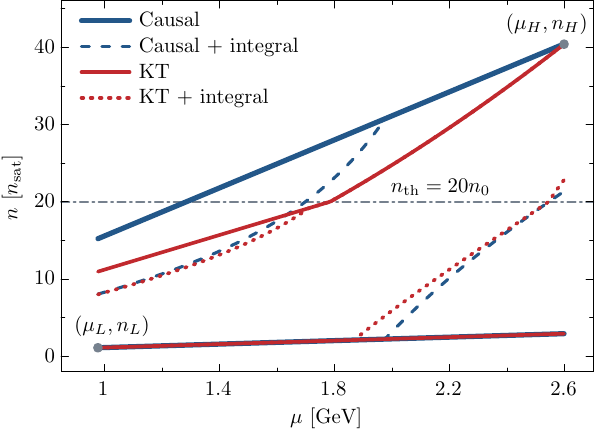}
    \caption{Bounds on the baryon density, $n(\mu)$, as a function of baryon chemical potential $\mu$, for the $\pqcd$ scale parameter $X=2$. Solid, blue lines represent bounds from causality, i.e., $\cs \leq 1$. Dashed, blue lines show the integral constraints (see Section~\ref{sec:causal} for details). Solid and dashed red lines show bounds from kinetic theory and kinetic theory combined with integral constraint, respectively (see Section~\ref{sec:kt} for details). We note that the integral constraints are not shown outside of the regions already constrained by either bounds. The gray dots mark the theoretical constraints from $\xeft$ at low density and pQCD at high density. The gray, dotted vertical line marks the density threshold $\nkt$ above which the kinetic theory constraint is applied. Here, as an example, $\nkt=20\,\nsat$ was used. Note that the lower causal and kinetic theory bounds overlap in this example.}
    \label{fig:n_u}
\end{figure}

For convenience, we define the stiffest EOS going through a point $(\mu_x, n_x)$ as
\begin{equation}
n_x(\mu) \equiv n(\mu; \mu_x, n_x) = \frac{n_x}{\mu_x} \mu \textrm .
\end{equation}
The lower and upper causal bounds, $n_L(\mu)$ and $n_H(\mu)$, correspond to the stiffest EOSs that pass through the points $(\mu_L, n_L)$ and $(\mu_H, n_H)$, respectively. This is demonstrated in Figure~\ref{fig:n_u}.

\begin{figure}[t!]
    \centering
    \includegraphics[width=1.0\linewidth]{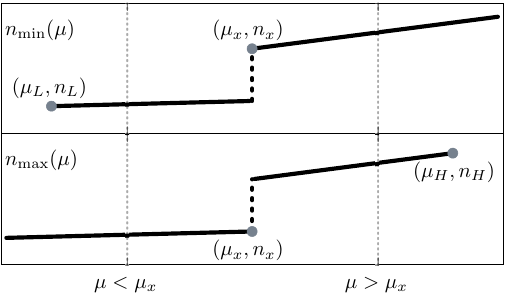}
    \caption{The construction of the density $n_{\rm min}(\mu)$ that maximizes the thermodynamic pressure at $\mu > \mu_x$ (top panel) and density $n_{\rm max}(\mu)$ that minimize the thermodynamic pressure at $\mu < \mu_x$ (bottom panel). Note that the functions $n_{\rm min}(\mu)$ and $n_{\rm max}(\mu)$ are extrapolated from $(\mu_L,n_L)$ and $(\mu_H, n_H)$, respectively.}
    \label{fig:n_u_schematic_cs1}
\end{figure}

The requirement that the pressure difference, $\Delta p$, between two chemical potentials, $\mu_L$ and $\mu_H$, is fixed imposes further (integral) constraints by fixing the area under $n(\mu)$ between them,
\begin{equation}\label{eq:dp_constraint}
    \Delta p  = \int\limits_{\mu_L}^{\mu_H} \dd \mu\; n\left(\mu\right) \textrm.
\end{equation}
By considering functions $n_{\rm min}(\mu)$ ($n_{\rm max}(\mu)$) that go through a fixed point $(\mu_x,n_x)$ and minimize (maximize) the pressure $\Delta p_{\rm max}$ ($\Delta p_{\rm min}$), one can rule out particular, otherwise allowed, regions in the $n(\mu)$ plane. These functions are given by
\begin{equation}
    n_{\rm min}(\mu) =
\begin{cases}\label{eq:n_min_casual}
n_L(\mu), & \mu < \mu_x \\[4pt]
n_x(\mu),  & \mu > \mu_x
\end{cases}\textrm,
\end{equation}
and
\begin{equation}
    n_{\rm max}(\mu) =
\begin{cases}\label{eq:n_max_casual}
n_x(\mu), & \mu < \mu_x \\[4pt]
n_H(\mu),  & \mu > \mu_x
\end{cases}\textrm.
\end{equation}
At $\mu_x$ the functions $n_{\rm min}(\mu)$ and $n_{\rm max}(\mu)$ feature first-order phase transitions from $n_L(\mu_x)$ to $n_x$ and from $n_x$ to $n_H(\mu_x)$, respectively. Their schematic construction is shown in Figure~\ref{fig:n_u_schematic_cs1}. The point $(\mu_x,~n_x)$ has to be determined separately for both $n_{\rm min}(\mu)$ and $n_{\rm max}(\mu)$ such that Eq.~\eqref{eq:dp_constraint} is met, i.e., $\Delta p_{\rm min} = \Delta p_{\rm max} = p_H - p_L$. This procedure yields both the upper and lower bounds on $n(\mu)$. The simultaneous causal and integral constraints are depicted in Figure~\ref{fig:n_u}. We note that the general conditions for the integral constraint can be solved analytically for the class of constant-speed-of-sound EOSs~\citep{Komoltsev:2021jzg}.

\begin{figure}[t]
    \centering
    \includegraphics[width=1.0\linewidth]{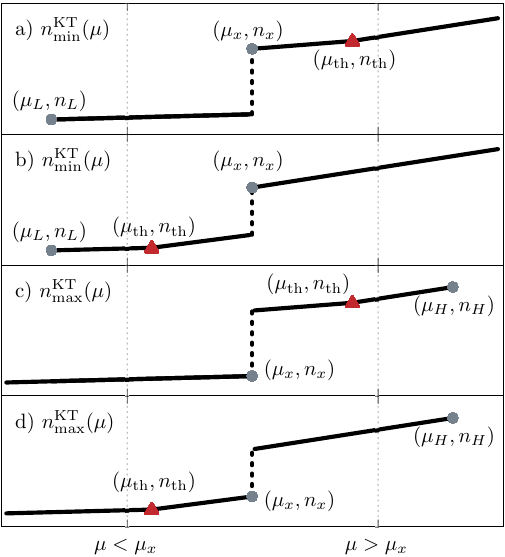}
    \caption{The construction of the density that includes the kinetic theory constraint at $n > \nkt$ (or equivalently at $\mu > \ukt$). The density $n^{\rm KT}_{\rm min}$ (panels {\it a} and {\it b}) maximizes the thermodynamic pressure at $\mu > \mu_x$ and $n^{\rm KT}_{\rm max}$ (panels {\it c} and {\it a}) minimizes the thermodynamic pressure $\mu < \mu_x$. The construction of the densities depends whether $\ukt > \mu_x$ (panels {\it a} and {\it c}) or $\ukt < \mu_x$ (panels {\it b} and {\it d}). The red triangles mark the onset of the kinetic theory ansatz. Note that the functions $n^{\rm KT}_{\rm min}$ and $n^{\rm KT}_{\rm max}$ are extrapolated from $(\mu_L,n_L)$ and $(\mu_H, n_H)$, respectively.}
    \label{fig:n_u_schematic_kt}
\end{figure}

\section{Bounds from kinetic theory}
\label{sec:kt}

The KT bound in Eq.~\eqref{eq:cs2_kinetic_eq} can be expressed as a differential equation for the stiffest EOS admissible by the constraint using the zero-temperature expression for the thermodynamic pressure $p = n^2 \dd \left(\eps/n\right) / \dd n$,
\begin{equation}\label{eq:kt_diff_eq}
    n^2\frac{\dd^2\eps(n)}{\dd n^2} + \frac{n}{3}\frac{\dd \eps(n)}{\dd n} - \frac{4}{3}\eps(n) = 0 \textrm.
\end{equation}
Solving this equation yields
\begin{equation}\label{eq:kt_eps}
\begin{split}
    \eps(n) &= \alpha_+ n^{\beta_+} + \alpha_- n^{\beta_-} \textrm, \\
\end{split}
\end{equation}
where $\beta_\pm = \left(1\pm\sqrt{13}\right)/3$ and $\alpha_\pm$ are integration constants. The pressure $p(n)$ and chemical potential $\mu(n)$ can be obtained through the zero-temperature Gibbs-Duhem equation $p+\eps =\mu n$ and $\mu = \dd \eps / \dd n$.

Since the microscopic regime where the relativistic kinetic-theory description of dense QCD matter becomes valid remains uncertain, we introduce the onset density $\nkt$ as a free parameter with $n(\mu_L) < \nkt < n(\mu_H)$. It marks the transition above which we impose the KT constraint on the speed of sound. Varying $\nkt$ allows us to quantify how early the KT regime must set in to significantly affect the stiffness of the EOS. At $n < \nkt$, the stiffest allowed EOS remains luminal, i.e., $c_s^2=1$, while at $n >\nkt$, the KT constraint sets in and from Eq.~\eqref{eq:cs2_kinetic_eq} one gets that
\begin{equation}
    c_s^2 = \frac{n}{\mu}\frac{\dd \mu}{\dd n}\leq c_{s, \rm KT}^2 \;\Longrightarrow \;\frac{\dd n}{\dd \mu} \geq \frac{n}{\mu}\left(1-\frac{4}{3}\frac{p}{\mu n}\right)^{-1} \textrm,
\end{equation}
where the zero-temperature Gibbs-Duhem equation was used. The bracket is always smaller than unity; thus, the slope of the stiffest EOS allowed by kinetic theory is always larger than that of luminal EOS. This way, the maximum speed of sound allowed by kinetic theory further constrains the allowed $n(\mu)$ space. 

\begin{figure}[t]
    \centering
    \includegraphics[width=1.0\linewidth]{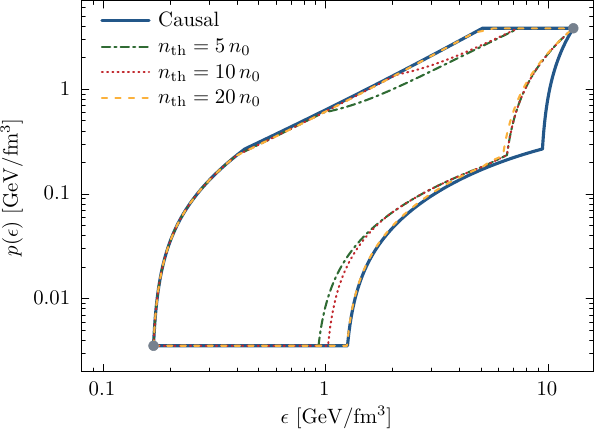}
    \caption{Constraints on the thermodynamic pressure, $p$, as function of energy density, $\eps$, obtained by mapping the allowed $n(\mu)$ space for different values of $\nkt$ (see text for details). The gray circles mark the theoretical constraints from $\chi$EFT at low density and pQCD at high density.}
    \label{fig:p_e}
\end{figure}

The construction of the constraints on $n(\mu)$ is similar to the case discussed in Section~\ref{sec:causal}. However, once the density reaches the threshold value, $\nkt$, the EOS has to switch to that of the kinetic theory,
\begin{equation}\label{eq:kt_caus_lh}
    n_{\rm L/H}^{\rm KT}(\mu) =
\begin{cases}
n_{{\rm L}/x}(\mu), & \mu \le \ukt \\
n_{\rm KT}(\mu),  & \mu > \ukt 
\end{cases}\textrm.
\end{equation}
Because $\nkt>n_L$, the lower bound on density, $n_{\rm L}^{\rm KT}(\mu)$, starts as the luminal EOS, $n_L(\mu)$, and switches to the KT EOS at $\ukt$ such that $n_L(\ukt) = \nkt$. The corresponding upper bound on density, $n_{\rm H}^{\rm KT}(\mu)$, is obtained by assuming that the EOS jumps from $n_L$ to $n_x$ at $\mu_L$. The density $n_x$ has to be chosen so that the upper bound goes through the high-density constraint at $\mu_H$, i.e., $n_{\rm H}^{\rm KT}(\mu_H) = n_H$. We remark that if $n_x(\mu_L) > \nkt$, the upper bound $n_{\rm H}^{\rm KT}(\mu)$ would be given entirely by the KT EOS fixed at $n_L$. 

The function $n_{\rm KT}(\mu)$ is given as a solution of the Eq.~\eqref{eq:kt_diff_eq}. However, to fully determine the KT EOS, one needs to know the pressure, $p(\ukt)$ at $\ukt$, which is obtained by integrating the density below $\ukt$.

We demonstrate the effect of the KT constraint for $\nkt = 20\,\nsat$ in Figure~\ref{fig:n_u}. The upper bound reduces significantly compared to the causal bound. On the other hand, the lower bound remains unchanged. This is because the lower causal bound $n_L(\mu_H) \approx 3\,\nsat < \nkt$. Note that as $\nkt$ approaches $n_H$, the impact of the KT constraint becomes ineffective and the bound smoothly reduces to the causal one.

With the kinetic-theory constraint, the construction of the integral constraint must take into account the KT onset density $\nkt$. The density functions that minimize and maximize the thermodynamic pressure, $n_{\rm min}^{\rm KT}$ and $n_{\rm max}^{\rm KT}$, respectively, have to take into account two cases. First, when $n_x < \nkt$, for which the densities are given as
\begin{equation}
    n_{\rm min}^{\rm KT}(\mu) =
    \begin{cases}\label{eq:n_min_casual_kt1}
   n_L(\mu), & \mu < \mu_x \\
    n_x(\mu),  & \mu_x < \mu \le \ukt \\
    n_{\rm KT}(\mu),          & \ukt \le \mu
    \end{cases}\textrm.
\end{equation}
and
\begin{equation}
    n_{\rm max}^{\rm KT}(\mu) =
    \begin{cases}\label{eq:n_min_casual_kt2}
   n_x(\mu), & \mu < \mu_x \\
    n_{\rm th}(\mu),  & \mu_x < \mu \le \ukt \\
    n_{\rm KT}(\mu),          & \ukt \le \mu
    \end{cases}\textrm,
\end{equation}
and second, when $n_x > \nkt$, for which the densities read
\begin{equation}
    n_{\rm min}^{\rm KT}(\mu) =
    \begin{cases}\label{eq:n_min_casual_kt3}
    n_L(\mu),              & \mu < \ukt \\
    n_{\rm KT}^{(1)}(\mu), & \ukt < \mu < \mu_x \\
    n_{\rm KT}^{(2)}(\mu), & \mu_x < \mu
    \end{cases}\textrm,
\end{equation}
\begin{equation}
    n_{\rm max}^{\rm KT}(\mu) =
    \begin{cases}\label{eq:n_min_casual_kt4}
    n_{\rm th}(\mu),             & \mu < \ukt \\
    n_{\rm KT}^{(1)}(\mu), & \ukt < \mu < \mu_x \\
    n_{\rm KT}^{(2)}(\mu), & \mu_x < \mu
    \end{cases}\textrm.
\end{equation}
The density $\nkt(\mu)$ is fixed at $\ukt$, i.e., $\nkt(\ukt) = n_x(\ukt)$ and $\pkt(\ukt) = p_x(\ukt)$. Schematic construction of the functions $n_{\rm min}^{\rm KT}(\mu)$ and $n_{\rm max}^{\rm KT}(\mu)$ is shown in Figure~\ref{fig:n_u_schematic_kt}.

\begin{figure}[t]
    \centering
    \includegraphics[width=1.0\linewidth]{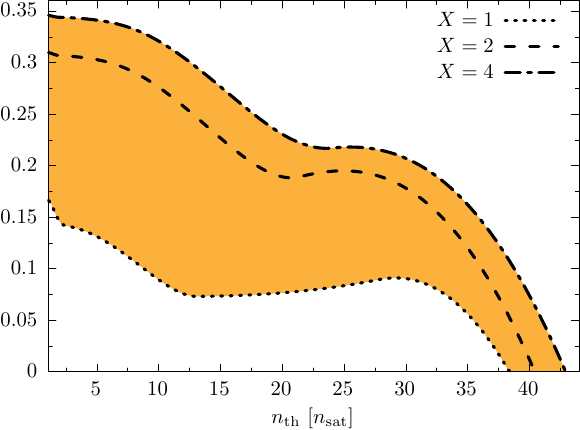}
    \caption{Fraction of the area excluded in the $p-\eps$ plane as a function of the onset density for the kinetic-theory constraint, $\nkt$,  in the units of saturation density. The effect of varying the scale parameter $X \in [1, 4]$ is included in the orange band.}
    \label{fig:area}
\end{figure}

The effect of imposing the integral constraint is shown in Figure~\ref{fig:n_u} for $\nkt = 20\,\nsat$ as an example. The inclusion of the KT bound markedly reduces the allowed range of densities at a given chemical potential. The upper branch becomes significantly softer once the KT constraint is applied above the threshold, while the lower branch remains essentially unchanged. As a result, the admissible region in the $n(\mu)$ plane contracts, anticipating a correspondingly narrower $\eps-p$ band.

\section{Kinetic-theory bound on \texorpdfstring{$\eps$-$p$}{}}

The constraint obtained in the $\mu-n$ plane can be mapped to the $\eps-p$ plane using the procedure outlined in Refs.~\cite{Komoltsev:2021jzg, Fujimoto:2023unl}. Figure~\ref{fig:p_e} shows the allowed range of $p(\eps)$ for different values of $\nkt$. The combined effect of the KT and integral constraints appears as a narrowing of the admissible EOS region. We note that because the KT condition turns on only at higher densities (above $\nkt$), its influence is concentrated in the high-density part of the EOS.

Figure~\ref{fig:area} shows the fraction of excluded area in the $\eps-p$ plane compared to the band allowed by causality. As $\nkt$ increases, the fraction of excluded area decreases. Introducing the KT constraint at densities close to $n_{\rm pQCD}$ does not offer a substantial reduction in the allowed $\eps-p$ space. However, even for moderate thresholds such as $\nkt \simeq 25\,\nsat$, the excluded fraction already reaches $10-23\%$ as the scale parameter $X$ is varied between $1$ and $4$. We note that larger values of $X$ lead to a correspondingly smaller excluded fraction. The largest excluded area is obtained for the smallest $\nkt$ and goes up to $35\%$ for $\nkt =1.1\,\nsat$ for $X=4$. However, we note that such a small $\nkt$ is probably unrealistic.

\section{Conclusion}

We have explored how relativistic kinetic theory (KT) constrains the stiffness of dense QCD matter. Building upon the model-agnostic construction of the equation of state (EOS) bounded by chiral effective field theory at low densities and perturbative QCD at high densities, we imposed an additional constraint on the speed of sound. This bound follows from the positivity of the relaxation time and the stability of the kinetic-theory dispersion relation, and it becomes effective only above a threshold density at which a quasi-particle description is assumed to apply. For any chosen value of onset density, imposing the KT condition removes a significant subset of causal and thermodynamically admissible interpolations, yielding a narrower admissible EOS band. Smaller values of the onset density produce stronger restrictions, while for larger values the effect smoothly reduces to the purely causal construction. In this way, kinetic theory provides a physically motivated, parameter-dependent upper bound on the stiffness of dense matter.

Our results indicate that transport-theory considerations can provide a physically motivated way of limiting the high-density behavior of the EOS. While the practical implications for neutron stars depend on the onset density of KT applicability, the framework offers a useful theoretical prior for connecting microscopic QCD dynamics with astrophysical constraints.

\medskip
\acknowledgments
The author thanks Pasi Huovinen for constructive remarks and Aleksandra Niewczas for helpful comments on the presentation of data. This work was partly supported by the program \textit{Excellence Initiative–Research University} of the University of Wroc\l{}aw, funded by the Ministry of Education and Science.

\bibliography{ref_biblio}
\end{document}